# Flexible generation/conversion/exchange of fiber-guided orbital angular momentum modes using helical gratings

LIANG FANG AND JIAN WANG*

*Wuhan National Laboratory for Optoelectronics, School of Optical and Electronic Information, Huazhong University of Science and Technology, Wuhan 430074, Hubei, China*
**Corresponding author: jwang@hust.edu.cn*

**By exploiting helical gratings (HGs), we propose and simulate flexible generation, conversion and exchange of fiber-guided orbital angular momentum (OAM) modes. HGs can enable the generation of OAM modes, and the OAM conversion between two arbitrary modes guided in fibers. A specific HG can exchange the OAM states of a couple of OAM modes, i.e, OAM exchange. In addition, a Fabry-Perot (F-P) cavity cascaded with two identical reflective HGs can reflect converted OAM modes with a comb spectrum. The HGs-based generation / conversion / exchange of OAM modes are dependent on helix period, orientation, and the fold number of helical fringes. The proposed method of generation, conversion, and exchange of fiber-guided OAM modes using HGs is flexible and well compatible with OAM fibers, featuring a high conversion efficiency close to 100% and a conversion bandwidth about 10 nm in transmission spectra, while less than 1 nm in reflection spectra.** 

Recently, immense attentions have been increasingly attracted on mode-division multiplexing, especially the multiplexing technology with orbital angular momentum (OAM) modes as the linear combination of even and odd vector modes with a $\pi/2$ phase shift [1,2] in special designed fibers, such as vortex fiber or ring-core fiber. These fibers are available to lift the degeneracy between vector modes that may tend to degenerate into linearly polarized (LP) mode in conventional fiber, and hence support stable OAM modes with little modal crosstalk, which have been used in transmitting OAM modes [2-5]. However, incident OAM modes are usually excited by free-space OAM beams generated outside fibers with bulky devices such as spatial light modulators (SLMs) [2]. So a more compact, compatible and high efficient method of generating OAM modes directly in fibers has become a key challenge [6, 7]. Proposed schemes on the generation of OAM modes in fibers include the conventional realizations by microbend gratings with a metallic block with grooves [8], the all-fiber methods using special ring fiber structures [9], helical core fiber [10], chiral fiber gratings [7], and helical fiber Bragg gratings [11]. Conventional optical fiber gratings have developed into a maturational technology with wide applications in optical communications and fiber sensing in recent decades [12, 13]. As for helical gratings (HGs), it would be a challenge to fabricate it, but perhaps could be done by rotating the fiber when writing them by point-by-point method or a phase mask in about one micrometer range [11, 14, 15].

In this Letter, we comprehensively analyze and study all the cases of OAM mode couplings with multiple-fold HGs inscribed in ring-core fibers. The analyses and simulations are based on OAM mode-coupled theory. Our presentations contain the types of reflective and transmissive HGs, as well as their influence on spin angular momentum (SAM) of converted OAM modes. We also show an interesting function of both OAM and power exchange between two co-directional propagating modes using a specific HG. Additionally, we exhibit the ability of multi-channel conversion of OAM modes using two identical reflective HGs cascaded as a Fabry-Perot (F-P) cavity.

The modulation function of a $|l|$-fold HG inscribed in ring-core fiber can be simply expressed by

$$\delta n(r,\phi,z) = \Delta n \rho(r) \cos\left[ l\left(\phi + \sigma \frac{2\pi}{\Lambda} z\right)\right] \quad (a_1 \le r \le a_2), \tag{1}$$

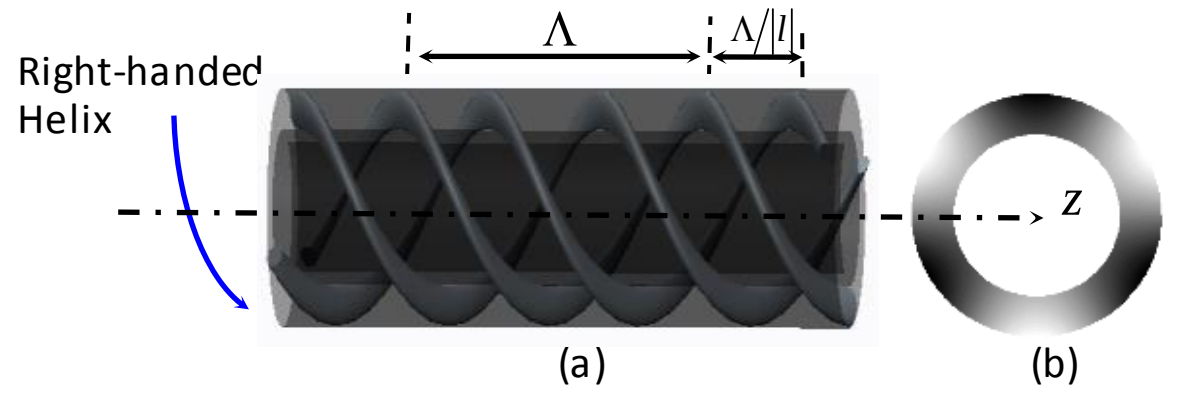

**Fig. 1.** (a) Longitudinal and (b) transverse modulation patterns of a 3-fold right-handed HG$\langle -1,3\rangle$.

where $\Delta n$ is the modulation strength, the absolute value of $l$ indicates the fold number of HG's helical fringes, $\Lambda$ denotes the period of HG's single fringe, $\sigma$ the helix orientation, $\sigma = +1$ and $-1$ correspond to left-handed and right-handed helix, respectively, $\rho(r)$ determinates the saturability of index modulation in the cross section defined by $\rho(r) = r^2/a_2^2$, and $a_1$ and $a_2$ correspond to inside and outside radii of the structure of ring-core waveguide. In this Letter, we use the sign of $\langle\sigma,|l|\rangle$ to indicate the character of HGs. As shown in Fig. 1, a 3-fold right-handed HG is inscribed in the waveguide layer of a ring-core fiber, and its longitudinal and transverse refractive index modulation patterns correspond to Fig. 1(a) and 1(b), respectively. Note that for simplicity the center and cladding structure of this ring-core fiber are not shown here.

The transverse electric field of a circularly polarized OAM mode denoted with $|s_n,n\rangle$ as reference [10] in optical fiber can be expressed with a matrix,

$$\mathbf{E}_n^t(r,\phi,z)=F_n(r)e^{i(n\phi-\beta_n z)}\begin{bmatrix}1\\ s_n i\end{bmatrix}, \quad \textbf{(2)}$$

where $n$ and $s_n$ denote the topological charge of OAM mode and the charge of SAM, respectively, $s_n=-1$, $+1$ or $0$ correspond to right circular, left circular or linear polarization, respectively, $\omega$ is angular frequency, $\beta_n$ represents the propagation constant of the OAM mode, defined by $\beta_n=2\pi n_{eff}/\lambda$, $n_{eff}$ effective refractive index of corresponding vector mode in fibers [5], and $F_n(r)$ corresponds to radial function of electric field.

When an OAM mode $|s_n,n\rangle$ with amplitude A is perturbed by a $|l|$-fold HG, it may couple its power into another OAM mode $|s_m,m\rangle$ with amplitude B. The coupled-mode equations based on OAM modes can be deduced and written as [16, 17]

$$\begin{cases}\dfrac{dA}{dz}+i\kappa_{nm}B\,e^{-i2\delta z}=0\\ \dfrac{dB}{dz}+i\kappa_{mn}^{*}A\,e^{i2\delta z}=0\end{cases}, \quad \textbf{(3)}$$

where $\kappa_{nm}=-\kappa_{mn}^{*}$, and

$$\kappa_{nm}=\frac{\omega\varepsilon_0 n_2^2\Delta n(1+s_n s_m)}{4a_2^2}\int_{a_1}^{a_2}\int_0^{2\pi}F_n(r)^{*}\cdot F_m(r)\exp\left[i(m-n+l)\phi\right]r^3 dr d\phi$$
$$\propto(1+s_n s_m)\cdot\int_0^{2\pi}\exp\left[i(m-n+l)\phi\right]d\phi \quad \textbf{(4)}$$

where $*$ denotes operation of conjugate transpose, $\omega$ and $\varepsilon_0$ are angular frequency and dielectric constant in vacuum, respectively, and $n_2$ is the refractive index of the ring-core waveguide. These differential equations can be solved in accordance with reference [17].

In the case of power coupling between two co-directional propagating OAM modes, the detuning factor is

$$\delta=\frac{1}{2}\left(\beta_m-\beta_n-\sigma l\frac{2\pi}{\Lambda}\right). \quad \textbf{(5)}$$

As for the type of co-directional coupling of HGs $\langle\sigma,|l|\rangle$, the rule of coupling OAM modes must be followed by:

1) The HGs do not change the polarization state of coupled OAM modes, due to the same signs of $s_n$ and $s_m$ that should be taken, as shown in Eq. (4), otherwise the coupling coefficient will become zero;

2) The topological charges $n$ and $m$ of two coupled OAM modes and the fold number of helical fringes $|l|$ should meet the angular momentum matching condition: $n-m=l$;

3) The phase matching condition takes the form: $\beta_m-\beta_n=2\pi\sigma l/\Lambda$, and the helix orientation is determined by $\sigma=\mathrm{sign}\left[(\beta_m-\beta_n)l\right]$.

We take the case of conversion between two right circularly polarized OAM modes as examples, the coupled OAM modes and HGs with respect to helical orientation and the fold number of helical fringes should be unified as follows:

$$|-1,-j\rangle\xleftrightarrow{\langle+1,|k-j|\rangle}|-1,-k\rangle\quad(j\neq k), \quad \textbf{(6-a)}$$

$$|-1,+j\rangle\xleftrightarrow{\langle+1,|j-k|\rangle}|-1,+k\rangle\quad(j\neq k), \quad \textbf{(6-b)}$$

$$|-1,-j\rangle\xleftrightarrow{\langle-1,j+k\rangle}|-1,+k\rangle\quad(j-k\geq 1), \quad \textbf{(6-c)}$$

$$|-1,-j\rangle\xleftrightarrow{\langle+1,j+k\rangle}\langle-1,+k\rangle\quad(j-k<1), \quad \textbf{(6-d)}$$

where the $j$ and $k$ are non-negative integers. The case of conversion between two left circularly polarized OAM modes is similar to the above analysis.

If the mode coupling occurs between two contra-directional propagating OAM modes, the detuning factor of the coupled mode equation has the form

$$\delta=-\frac{1}{2}\left(\beta_m+\beta_n+\sigma l\frac{2\pi}{\Lambda}\right). \quad \textbf{(7)}$$

Relative to the forward propagation, the reflected OAM mode would change its polarization direction, i.e., the sign of SAM will be reversed, as well as the charge sign of OAM due to reflection by HGs. It can be explained by that the relative phase between the forward and backward propagating modes is shifted by 180°, as do mirrors in free space propagation [18]. Therefore, the angular momentum matching fcondition should be replaced by $n+m=l$, and the phase matching condition is taken in the form of $\beta_m+\beta_n=-2\pi\sigma l/\Lambda$. The helical orientation is then determined by $\sigma=\mathrm{sign}\left[-(\beta_m+\beta_n)l\right]$.

Similarly, taking the case of conversion between one right and one left circularly polarized OAM modes as examples, the coupled OAM modes and corresponding HGs should be unified as follows:

$$|-1,-j\rangle\xleftrightarrow{\langle+1,j-k\rangle}|+1,+k\rangle\quad(j>k), \quad \textbf{(8-a)}$$

$$|-1,-j\rangle\xleftrightarrow{\langle-1,k-j\rangle}|+1,+k\rangle\quad(j<k), \quad \textbf{(8-b)}$$

$$|-1,+j\rangle\xleftrightarrow{\langle-1,j+k\rangle}|+1,+k\rangle\quad(j\neq k), \quad \textbf{(8-c)}$$

$$|-1,-j\rangle\xleftrightarrow{\langle+1,j+k\rangle}|+1,-k\rangle\quad(j\neq k), \quad \textbf{(8-d)}$$

$$|-1,-j\rangle\xleftrightarrow{uniform}|+1,+j\rangle, \quad \textbf{(8-e)}$$

where $j$ and $k$ are non-negative integer. Especially, for the OAM mode coupling between $|s,n\rangle$ and $|-s,-n\rangle$, the gratings needn't to be taken as helical modulation, a conventional uniform fiber Bragg gratings (FBGs) just can realize the OAM mode's reflection [18].

In the following simulations, the parameters of ring-core fiber are chosen as follows: the inside and outside radii of ring-core waveguide are $a_1=3.00$ μm, $a_2=4.40$ μm, respectively, and the refractive index of innermost layer and equivalently infinite cladding are set as $n_1=n_3=1.62$, that of ring-core waveguide is $n_2=1.689$. This kind of high contrast index ring-core fiber has been used to theoretically support and transmit OAM modes [6, 19], and can be achieved in practical fabrication [20]. In this Letter, the refractive index difference is taken as $\Delta=0.04$ to support 12 vector modes with negligible degeneracy ( $\Delta n_{eff} > 10^{-4}$ ), corresponding to enough OAM modes supported with low inter-mode crosstalk at 1550nm. In simulations, to clarify the total angular momentum of mode, we use helical wavefronts to represent OAM states, and left or right-handed circles to denote the circular polarization states of OAM modes.

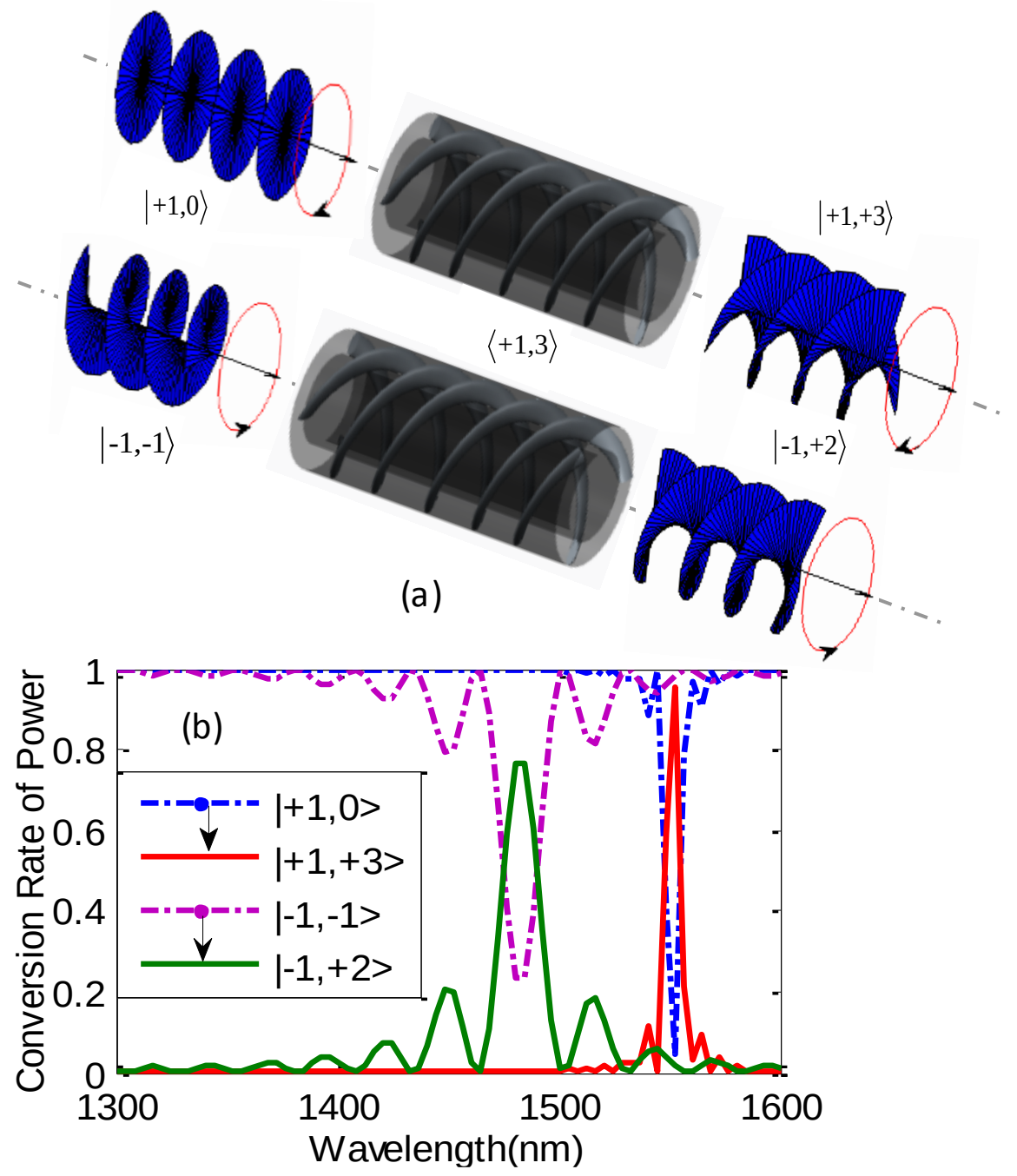

**Fig. 2**. The conversion of modes from $|+1,0\rangle$ to $|+1,+3\rangle$, and from $|-1,-1\rangle$ to $|-1,+2\rangle$ by HG $\langle +1,3\rangle$. (a) Conversion of helical phase. (b) Conversion spectrum of optical power.

Firstly, we show the case of co-directional OAM mode coupling. We design a 3-fold HG to show the conversion of modes from $|+1,0\rangle$ to $|+1,+3\rangle$ and from $|-1,-1\rangle$ to $|-1,+2\rangle$, respectively. The grating parameters are designed with helical character $\langle +1,3\rangle$, grating period $\Lambda$=362.0 μm, length $L = 2.0$ cm, and modulation strength $\Delta n = 4.83\times10^{-5}$, respectively, based on the conditions of both phase matching and angular momentum matching, as well as strong coupling interaction [17]. The phase conversion and power exchange of these modes is shown in Fig. 2. From Fig. 2(b), one can see that there are two resonant wavelengths at 1465 nm and 1550 nm respectively corresponding to two kinds of OAM conversion shown in Fig. 2(a). The coupling efficiency can be increased to close to 100% by increasing grating length or modulation strength. Actually, according to the detuning factor shown in Eq. (5), the phase matching condition of OAM modes coupling can be amended as $\beta_m - \beta_n = 2\pi\sigma l/t\Lambda$, $(t = 1,2,3,...,l)$, thus one specific HG can realize more than one type of OAM conversion. Here, for coupling from $|+1,0\rangle$ to $|+1,+3\rangle$, $t=1$; while for coupling from $|-1,-1\rangle$ to $|-1,+2\rangle$, $t=3$. The bandwidth of conversion from $|-1,-1\rangle$ to $|-1,+2\rangle$ is larger than that of conversion from $|+1,0\rangle$ to $|+1,+3\rangle$. Such phenomenon can be explained as follows. For the latter case, mode coupling interacts with more grating periods than that for the former case due to the smaller actual period of HG in the case of the latter coupling, in other words, the peak height increases with increasing the number of periods and modulation strength while the peak width decreases [17]. One can predict that this HG $\langle +1,3\rangle$ adds an additional OAM in the form of $\exp(3i\phi)$ to the coupled modes $|+1,0\rangle$ and $|-1,-1\rangle$, which can be explained by the expression of coupling coefficient in Eq. (4). We can achieve the conversion between two arbitrary OAM modes using the HGs at arbitrary wavelength, provided that both the phase matching and angular momentum matching conditions are well satisfied.

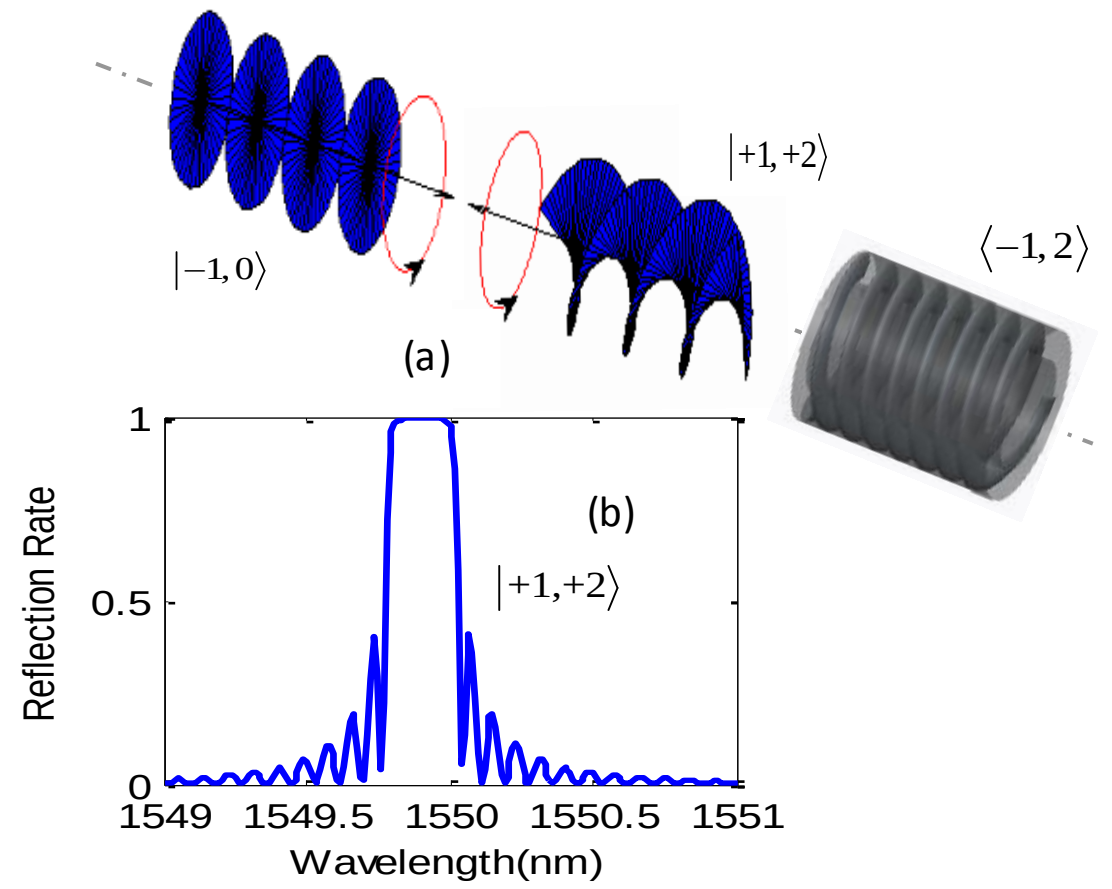

**Fig. 3**. The conversion from fundamental mode $|-1,0\rangle$ to OAM mode $|+1,+2\rangle$ by reflective HG $\langle -1,2\rangle$. (a) Conversion of phase front. (b) Reflection spectrum of optical power.

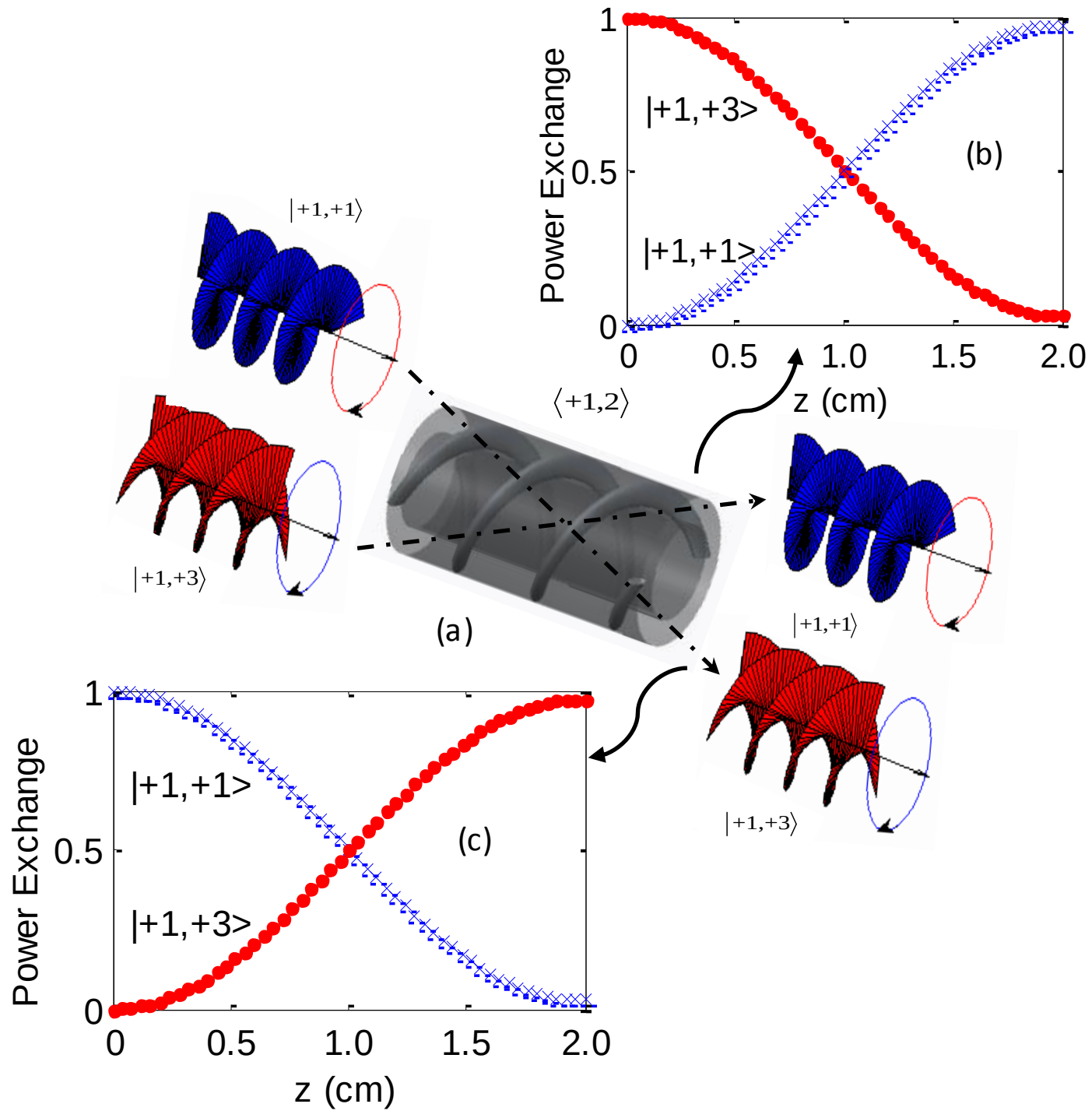

**Fig. 4**. OAM and power exchange between two co-directional propagating modes $|+1,+1\rangle$ and $|+1,+3\rangle$ through HG $\langle +1,2\rangle$. (a) Exchange of helical phase. (b) and (c) Exchange of optical power.

We design a reflective HG $\langle -1,2\rangle$ to generate OAM mode $|+1,+2\rangle$ contra-directionally from the fundamental mode $|-1,0\rangle$. The grating parameters are set as $\Lambda$=0.935 μm, $L = 0.80$ cm, and $\Delta n = 3.8\times10^{-4}$. The conversion of phase front and optical power of these modes are demonstrated in Fig. 3. One can see that the reflected bandwidth of contra-directional spectrum is less than 1 nm when compared with the large conversion bandwidth more than 10 nm in the co-directional coupling. This difference of bandwidth is cased by the resonance condition of gratings and wavelength dispersion [17]. It should be noted that the reflected OAM mode has change its polarization direction, i.e., the SAM has reverse its sign due to reflection by the HG.

In addition we present the function of OAM and power exchange between two co-directional propagating modes through one HG in Fig. 4. The grating parameters are set as $\Lambda = 273.0$ μm, $L = 2.0$ cm and $\Delta n = 4.5\times10^{-5}$. Two modes $|+1,+1\rangle$ and $|+1,+3\rangle$ are launched into the HG simultaneously. After transmitting through the HG $\langle +1,2\rangle$, the OAM states of these two modes are exchanged as shown in Fig. 4(a). Meanwhile, the power evolutions of two OAM modes along the length of HG are shown in Fig. 4(b) and 4(c). The efficiency of OAM exchange here is simulatively greater than 98%. In theory, the exchange efficiency can reach 100%, by adequately increasing the modulation strength or grating length. This coupling function of HGs may have a potential application in data exchange in OAM communications [1].

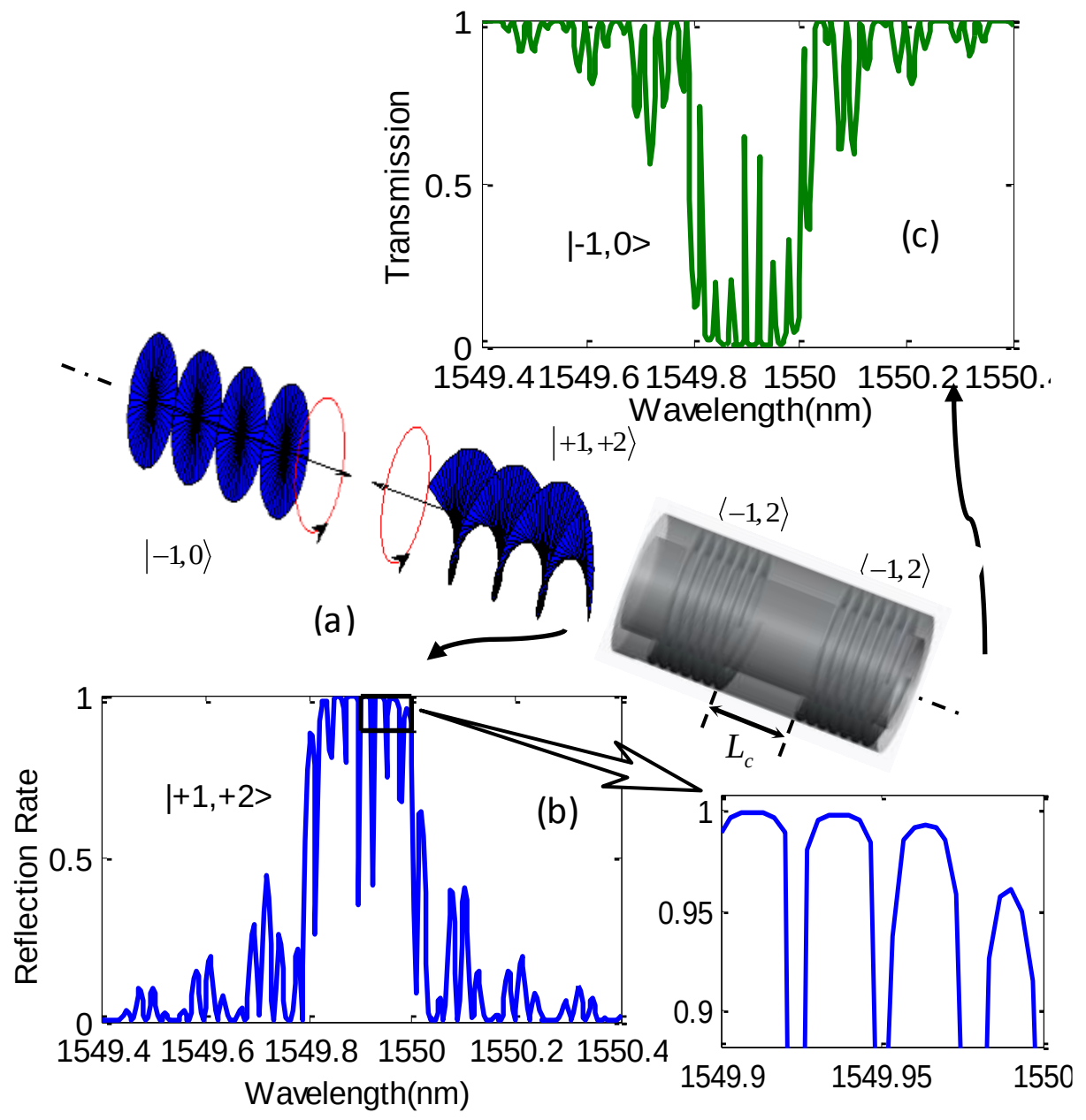

**Fig. 5**. (a) The schematic diagram of F-P cavity consisting of two identical reflective HGs $\langle -1,2\rangle$. (b) Reflection spectrum. (c) Transmission spectrum of F-P cavity.

Finally, we study the OAM spectral character of two identical reflective HGs cascaded as an F-P cavity, similar to that consisting of two FBGs [21]. The schematic diagram of this F-P cavity is shown in Fig. 5(a). The simulated reflection and transmission spectra are shown in Fig. 5(b) and 5(c), respectively. The grating parameters set are the same with those in Fig.(3) , and the length of F-P cavity is $L_c = 2.0$ cm. One can see that the angular momentum state of transmission spectrum is $|-1,0\rangle$, while that of reflection spectrum is $|+1,+2\rangle$, which is due to the back and forth conversion of the angular momentum states by two reflective identical HGs. Since the spectra in Fig. 5 are superpositions of the single HG reflection (or transmission) spectrum and the longitudinal modes of the F-P cavity, they appear as comb spectra [22]. The reflection efficiency of HG-based F-P cavity can be deduced as [21]

$$R_{F-P} = \frac{4R\sin^2\left[(\beta_n+\beta_m)L_c/2+\delta L\right]}{(1-R^2)+4R\sin^2\left[(\beta_n+\beta_m)L_c/2+\delta L\right]}, \qquad (9)$$

where $R$ denotes the reflectivity of each HG, and $\delta$ is the phase change, defined as Eq. (7). The F-P cavity consisting of two identical reflective HGs has a special spectral property, and may be applied to spectral processing of OAM mode conversion.

In conclusion, we present a theoretical study of flexible generation, conversion, and exchange of fiber-guided OAM modes using helical optical gratings inscribed in ring-core fibers. We also show multi-channel conversion of OAM modes using a F-P cavity consisting of two identical reflective helical gratings. Theoretically, we can achieve arbitrary conversion between two fiber-guided OAM modes using well designed HGs. We believe that it is a promising method for fiber-based generation, date exchange, and spectra processing of OAM modes, and may have some potential applications in OAM multiplexing in fiber communications.

**Funding.** National Basic Research Program of China (973 Program, 2014CB340004); National Natural Science Foundation of China (NSFC, 11274131, 61222502 and L1222026); the Program for New Century Excellent Talents in University (NCET-11-0182); Wuhan Science and Technology Plan Project (2014070404010201); Fundamental Research Funds for the Central Universities (HUST, 2012YQ008 and 2013ZZGH003); the seed project of Wuhan National Laboratory for Optoelectronics (WNLO).